\documentclass{article}
\usepackage{arxiv}
\usepackage[utf8]{inputenc} % allow utf-8 input
\usepackage[T1]{fontenc}    % use 8-bit T1 fonts
\usepackage{hyperref}       % hyperlinks
\usepackage{url}            % simple URL typesetting
\usepackage{booktabs}       % professional-quality tables
\usepackage{amsfonts}       % blackboard math symbols
\usepackage{nicefrac}       % compact symbols for 1/2, etc.
\usepackage{microtype}      % microtypography
\usepackage{lipsum}		% Can be removed after putting your text content
\usepackage{graphicx}
\usepackage[numbers]{natbib}
\usepackage{doi}
\usepackage{amsmath,amssymb,geometry}
\usepackage{bm}
\usepackage{hyperref}

\title{A note on the shear-forced dynamics of tornadoes}

\author{\href{https://orcid.org/0000-0001-8947-8534}{\includegraphics[scale=0.07]{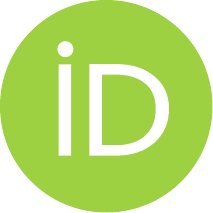}\hspace{1mm}Chanh Kieu}\thanks{Corresponding Author: ckieu@iu.edu. Current Affiliation: Center for Climate Research Singapore, Department of Weather Research, Core Modeling Development team, Singapore.} \\
	Department of Earth and Atmospheric Sciences\\
	Indiana University Bloomington, 47405, Indiana\\
}

\renewcommand{\shorttitle}{\textit{arXiv} Shear-forced tornado dynamics}

\hypersetup{
pdftitle={A template for the arxiv style},
pdfsubject={q-bio.NC, q-bio.QM},
pdfauthor={Chanh Kieu},
pdfkeywords={First keyword, Second keyword, More},
}

\begin{document}
\maketitle

\begin{abstract}
This note presents an axisymmetric model of a tornado-like vortex with internal vertical wind shear. By employing a prescribed incompressible circulation-strain flow that captures the horizontal variation of the vertical wind, we show that the model admits an exact viscous Feynman-Kac representation for the tornado structure. In the presence of vortex vertical shear, the tilting term induces radial phase mixing that rapidly deforms the vortex structure, thus causing the radius of the maximum azimuthal wind to broaden and shift upward. At the long-time limit, the vortex approaches the Burgers radial scale while both the wind and vorticity decay exponentially. This class of solutions may help explain why tornadoes often weaken rapidly once an internal vertical shear structure emerges after touchdown.
\end{abstract}

% keywords can be removed
\tableofcontents
\newpage

\section{Introduction}
Tornadoes are among the most extreme phenomena in the atmosphere. From a risk perspective, they are very destructive, capable of causing severe damages along their paths. From a scientific standpoint, tornadoes display however a beautiful manifestation of atmospheric fluid dynamics, exhibiting highly organized yet complex vortical structures.

Our increasing understanding of tornadoes has been developed primarily through laboratory experiments, field observations, Doppler radar measurements, and numerical simulations \citep{Rotunno1979, BluesteinGolden1993, Fiedler1994thermodynamic, RotunnoBluestein2024recent}. These studies have provided valuable insights into various aspects of tornado dynamics, internal structure, formation, evolution, and climatological characteristics. Together, they have established a comprehensive physical picture of tornado behavior over a wide range of spatial and temporal scales.

Despite these advances, significant challenges remain. From the observational perspective, tornadoes are relatively rare, short-lived, and spatially localized, making detailed observations difficult to obtain \citep{Wurman_etal2013finescale}. Thus, direct measurements are often limited to radar observations, photographic records, and occasional in situ measurements. From the numerical standpoint, their small spatial scales and rapid evolution also make them difficult to be resolved explicitly in current numerical weather prediction models, even for very high-resolution nowcasting models. Because of this, tornado warnings are typically issued only minutes to, at most, a few hours in advance, mostly relying on radar signatures or short-range environmental conditions that are favorable for tornadogenesis probability \cite{Gallo_etal2016forecasting, Herman_etal2018}.

While observational and modelling approaches have greatly improved our understanding of tornado formation, structure, and dynamics, less attention has been given to theoretical models that admit exact or nearly exact analytical solutions. Since any tornado solutions are ultimately dictated by the Navier--Stokes equations, it is natural to ask whether we can see how these solution families look like in a closed-form that could capture essential features of tornado-like vortices beyond the lens of numerical simulations. Of course, such analytical solutions cannot reproduce every complexity of real tornadoes. However, they can isolate the fundamental mechanisms governing vortex structure and evolution or provide physical insight that is often difficult to obtain from observations or numerical simulations alone.

This note aims to examine a family of simplified analytical solutions that can describe tornado-like vortices directly from the governing equations. Our objective is not to construct a complete model of tornadoes, but rather to demonstrate how tornado-like structures can emerge as mathematically consistent solutions of the Navier--Stokes equations under appropriate forcing. Such solutions can highlight how the complex behavior observed in the atmosphere may arise from the Navier--Stokes equations under suitable conditions. 

The rest of this note is organized as follows. Section 2 reviews some existing theoretical models of vortex dynamics together with several classical solutions that provide the foundation for tornado models. Section 3 introduces a shear-forced framework for tornado dynamics and derives a family of analytical solutions. Section 4 summarizes our key findings and some implications and weaknesses in our model for understanding tornado dynamics.

\section{Theoretical framework}
\subsection{Governing equations}
Given that tornadoes are short lifetime at small scale near the surface, we will set up our theoretical framework for tornado solutions using the incompressible fluid as in previous studies (see, e.g., \cite{Rotunno1979, Fiedler1994thermodynamic, Nolan2005new}), which can be summarized as follows:
\begin{equation}\label{eq:momentum}
\frac{\partial \mathbf{u}}{\partial t} + (\mathbf{u} \cdot \nabla)\mathbf{u} = -\frac{1}{\rho}\nabla p + \mathbf{g} + \nu\nabla^2\mathbf{u} + F_z\mathbf{k}
\end{equation}
\begin{equation}\label{eq:continuity}
%\frac{1}{r} \frac{\partial(r u)}{\partial r} +\frac{\partial w}{\partial dz} = 0.
\nabla \cdot \mathbf{u} = 0
\end{equation}
%\begin{equation}\label{eq:thermo}
%\frac{\partial \theta}{\partial t} + \mathbf{u} \cdot %\nabla\theta = Q_\theta
%\end{equation}
where $\mathbf{u},p$ denotes the wind vectors and pressure, $\mathbf{g}$ gravity, $\nu$ kinematic viscosity, and $\rho$ air density. Here, we also assume the fact that strong buoyancy is represented by convective-scale forcing, which produces the strong updrafts in supercell storms in the form of force $F_z$ as in previous studies. Physically, this forcing $F_z$ is related to microphysics processes and latent heat released by condensation within supercell thunderstorms, which can be parameterized in terms of CAPE (see, e.g., \cite{Fiedler1994thermodynamic, NolanFarrell1999}). In this study, we will not focus on details of such thermodynamic processes, so that our study will only be on the dynamics aspect of tornadoes, similar to previous theoretical models for tornadic dynamics.   

Given these governing equations, it is common to work with the vorticity for theoretical purposes rather than directly with wind field. So, let $\mathbf{\omega} \equiv \nabla \times \mathbf{u}$
\begin{equation}\label{eq:vort}
\frac{D\mathbf{\omega}}{Dt} = (\mathbf{\omega} \cdot \nabla)\mathbf{u} - \mathbf{\omega}(\nabla \cdot \mathbf{u}) + \frac{1}{\rho^2}\nabla\rho \times \nabla p + \nu\nabla^2\mathbf{\omega}
\end{equation}
In the cylindrical coordinate $(r,\theta,t)$, the axisymmetry can be used so that there is no $\theta$-dependence. In this case, the continuity equation allows us to introduce a streamfunction $\psi(r,z,t)$:
\begin{equation}\label{eq:def_streamfunction}
u = -\frac{1}{r} \frac{\partial \psi}{\partial z}, \qquad v = \frac{\Gamma}{r}, \qquad w = \frac{1}{r} \frac{\partial \psi}{\partial r},
\end{equation}
where, by convention, the three wind components $(u,v,w)$ in the cylindrical coordinate represent the radial, azimuthal, and vertical motion in $(r,\theta,z)$ directions, respectively.  

The advective operator becomes
\begin{equation}\label{eq:jacobi}
u \partial_r + w \partial_z = \frac{1}{r} \left( \psi_r \partial_z - \psi_z \partial_r \right) = \frac{1}{r} J(\psi, \cdot),
\end{equation}
where
\begin{equation*}
    J(\psi, f) = \psi_r f_z - \psi_z f_r.
\end{equation*}
Using the angular momentum $\Gamma = vr$ variable (also known sometimes as a swirl variable in tornado studies), the azimuthal momentum equation can be now written as
\begin{equation}\label{eq:swirl}
\Gamma_t + \frac{1}{r} J(\psi, \Gamma) = \nu \left(\Gamma_{rr} - \frac{1}{r} \Gamma_r + \Gamma_{zz} \right).
\end{equation}
while the equation for the vertical component of the vorticity vector $\zeta$ now satisfies
\begin{equation}\label{eq:vort_z}
\frac{D\zeta}{Dt} = -\zeta(\nabla_h \cdot \mathbf{u}_h) + \left( \frac{1}{r}\frac{\partial w}{\partial \theta}\frac{\partial u}{\partial z} - \frac{\partial w}{\partial r}\frac{\partial v}{\partial z} \right)+ \frac{1}{\rho^2} \nabla_h\rho \times\nabla_hp + \nu \nabla^2 \zeta,
\end{equation}
where the subscript $h$ denotes horizontal components or derivatives. Equations (\ref{eq:momentum}), (\ref{eq:swirl}), and (\ref{eq:vort_z}) for the azimuthal velocity $v$, circulation $\Gamma$, and vertical vorticity $\zeta$, respectively, provide the foundation for the tornado solutions to be examined in the following sections. The choice among these equations often depends on the specific problem under consideration, particularly its symmetry, physical assumptions, and analytical tractability. Before deriving the fully time-dependent tornado solutions that incorporate the effects of vertical wind shear, we first review two classical axisymmetric vortex models for which exact analytical solutions exhibiting tornado-like structures have been obtained.

\subsection{Steady-state Burgers vortex model}
The simplest tornado-like model with a genuine closed-form exact solution is the Burgers vortex \cite{Burgers1948}. It is not a full tornado model in the sense that it requires some prescribed flow fields for the radial wind in advance. However, it captures the key idea that rotation intensifies because the flow is stretched vertically, while viscosity prevents the core from collapsing to zero radius.

This Burgers vortex model is best seen by starting from the incompressible vertical vorticity equation Eq. (\ref{eq:vort_z}) in cylindrical coordinates under the assumptions of a steady flow and axisymmetric. Then Eq. (\ref{eq:vort_z}) becomes 
\begin{equation}\label{eq:Burger}
u \frac{\partial \zeta}{\partial r} + w \frac{\partial \zeta}{\partial z} = \zeta \frac{\partial w}{\partial z} + \nu \left( \frac{\partial^2 \zeta}{\partial r^2} + \frac{1}{r} \frac{\partial \zeta}{\partial r} \right)
\end{equation}

Specific for the Burgers vortex, let's consider a strain wind field of the form
\begin{equation}\label{eq:wind_strain}
u = -\frac{a}{2}r, \qquad w = az
\end{equation}
with strain rate $a > 0$. This flow field is not arbitrary, but represents a free-slip inflow towards the center with the convergence rate of $-a$. To be consistent with the continuity equation, the vertical motion $w$ must then accelerate upward. This strain flow setting is often held true at the base of meso-scale convective systems, which sets up favorable environmental conditions for upright development of thunderstorms and tornadoes. Previous modeling studies in fact often capture or impose similar conditions of inflow and vertical motion during the development of tornadoes (see, e.g, \cite{Nolan2005new, Rotunno_etal2016, Giove_etal2025}), thus justifying our use of the Burger model for the steady state of tornadoes.    

The exact steady vorticity solution under this strain wind field can be then obtained as
\begin{equation}\label{eq:Burger_solution}
\zeta(r) = \frac{M a}{4\pi\nu} \exp\left( -\frac{a r^2}{4\nu} \right)
\end{equation}
from which the azimuthal wind $v(r)$ can be derived from the momentum equation as
\begin{equation*}
v(r) = \frac{M}{2\pi r} \left[ 1 - \exp\left( -\frac{a r^2}{4\nu} \right) \right]
\end{equation*}
where $M$ is an amplitude that can be also interpreted as the Kelvin circulation of a tornado at radius $R$.
%
% Figure: Burger solution
%
\begin{figure}[ht!]
\centering
\includegraphics[width=8cm]{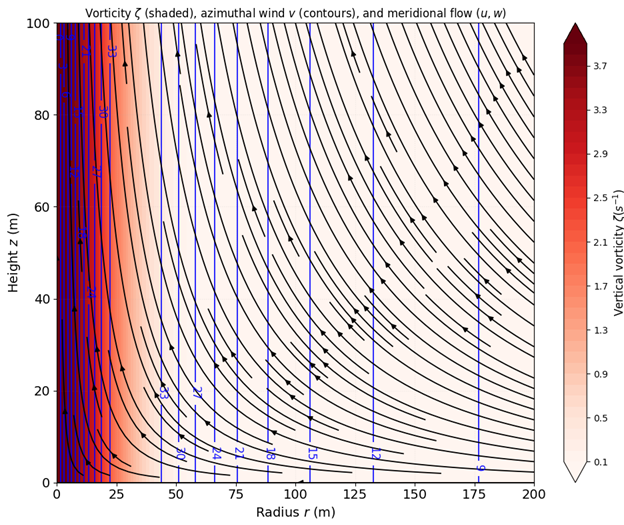}
\caption{An example of a radius-height cross section of the Burgers vortex vertical vorticity (shaded, unit s$^{-1}$ and the azimuthal wind amplitude (blue contours, unit ms$^{-1}$. Black solid contours denote the in-plane streamline of radial and vertical motion.} 
\label{fig:Burgers_vortex}
\end{figure}

This solution is of interest for the tornadic structure in several ways. Figure \ref{fig:Burgers_vortex} shows an example of the radial-height cross section of this solution. One notices clearly that the kinematic viscosity $\nu$ is critical to prevent the solution from collapsing. Consider a typical atmospheric condition for which $a \approx 10^{-1}\; s^{-1} , \nu = 50\; m^2s^{-1}$, the core width of tornadoes is about 45 $m$, which is reasonable in reality.  

Of course, the use of a prescribed strain flow immediately prevents us from applying this solution to far radii or high level above the PBL. In addition, the column structure as shown in Fig. \ref{fig:Burgers_vortex} instead of the funnel shape is clear evidence that the assumption of barotropic structure (i.e., no dependence on $z$) is generally not valid at all levels. However, the fact that the Burger equation, which is a part of the Navier-Stokes equations, could contain a solution of reasonable scales for tornadoes indicate that the existence of a tornadic steady-state solution is possible and permitted, so long as the strain wind are well maintained.   

\subsection{Kambe's time-dependent solution}
The exact solution (\ref{eq:Burger_solution}) is interesting as it captures some features of tornadoes, yet it is a steady-state solution. Thus, one may question wether this solution could be an asymptotic from some initial condition, and if so whether there exists any model to capture that asymptotic behavior. Surprisingly, there exists such a time-dependent model that indeed approaches the Burger steady-state solution, which was found by Kambe (1984 \cite{Kambe1984axisymmetric}). 

To set up for this solution, we assume again the same incompressible axisymmetric flow with a prescribed strain as in the Burger vortex model, but start from the full time-dependent equation for the vertical vorticity as follows: 
\begin{equation}\label{eq:Kambe_eqn}
\frac{\partial \zeta}{\partial t}
-\frac{a}{2}r\frac{\partial\zeta}{\partial r}
= a\zeta +\nu\left( \frac{\partial^2\zeta}{\partial r^2}
+\frac1r\frac{\partial\zeta}{\partial r} \right).
\end{equation}
Kambe used two coordinate variables $\xi,\tau$ defined as follows
\begin{equation}
\xi=re^{at/2}, \qquad \tau=\frac{e^{at}-1}{a}, \qquad
\zeta=e^{at}U(\xi,\tau),
\end{equation}
so that Eq. (\ref{eq:Kambe_eqn}) can be transformed into a diffusion equation of the form
\begin{equation}
\frac{\partial U}{\partial\tau} = \nu\left( \frac{\partial^2U}{\partial\xi^2} +\frac1\xi\frac{\partial U}{\partial\xi}\right).
\end{equation}

For any initial profile $\zeta(r,0) = \zeta(r)$, the solution is therefore the heat-kernel evolution in $\xi$-space. Upon transforming back to the $(r,t)$ coordinate, we obtain the solution in the integral form as follows:
\begin{equation}\label{eq:Kambe_general_sol}
\zeta(r,t) = e^{at} \int_{0}^{\infty} \frac{\eta \, d\eta}{2\nu\tau} \exp\left[ -\frac{\xi^2 + \eta^2}{4\nu\tau} \right] I_0\left( \frac{\xi\eta}{2\nu\tau} \right) \zeta(\eta),
\end{equation}
where $I_0$ the modified Bessel function. This is Kambe's general transient solution for the time-dependent Burger vortex model, which contains several interesting features.

To see the significance of this solution for our tornado problem, consider an initial vertical vorticity column that is in the form of a vortex filament, i.e.,
\begin{equation}
\zeta(r)=\Omega\delta(x)\delta(y)    
\end{equation}
which essentially represents an initial infinitely thin vorticity column (also known as a vortex rope) that is concentrated near a vortex axis. For this initial condition, the solution (\ref{eq:Kambe_general_sol}) can be obtained exactly as
\begin{equation}\label{eq:Kambe_delta_solution}
\zeta(r,t) = \frac{M a}{4\pi\nu(1 - e^{-at})} \exp\left[ -\frac{a r^2}{4\nu(1 - e^{-at})} \right].
\end{equation}
As $t \to \infty$, $1 - e^{-at} \to 1$ and as this solution can be seen to settle to the Burgers vortex solution (\ref{eq:Burger_solution}).

To bring more physical insight into this solution, we introduce a parameter "vortex core-width" $\sigma(t)$ as follows: 
\begin{equation}\label{eq:core_width}
\frac{d\sigma^2}{dt} = \nu - a\sigma^2 \rightarrow     \sigma^2(t) = \frac{\nu}{a}(1- e^{-at}),
\end{equation}
%(by construction, we set the integral constant to zero to be consistent with Eq. \ref{eq:Kambe_delta_solution}). 
Solution (\ref{eq:Kambe_delta_solution}) will then be re-written as
\begin{equation*}
\zeta(r,t) = \frac{M}{4\pi\sigma^2(t)} \exp\left( -\frac{r^2}{4\sigma^2(t)} \right).
\end{equation*}
At the limit $t \rightarrow \infty$, the core width relaxes exponentially to $\sigma_{\infty}^2 = \frac{\nu}{a}$, which is exactly the steady Burgers-vortex core width, and the corresponding Kambe's solution reduces to
\begin{equation}
\zeta_{\infty}(r) = \frac{M a}{4\pi\nu} \exp\left( -\frac{a r^2}{4\nu} \right).
\end{equation}
which is also the Burger's vortex solution (\ref{eq:Burger_solution}) as expected.

Using further the definition of $\zeta$ and the radial and vertical momentum equations, i.e.,
\[
\zeta = \frac{1}{r}\frac{\partial}{\partial r}(rv), \qquad
\frac{\partial p}{\partial r} = \rho \left( \frac{v^2}{r} - \frac{a^2}{4}r \right), \qquad
\frac{\partial p}{\partial z} =-\rho a^2 z
\]
we can derive the full solution for the azimuthal wind $v(r,t)$ and the pressure field $p(r,z,t)$ as: 
\begin{equation}\label{eq:Kambe_v_solution}
v(r,t) = \frac{M}{2\pi r} \left(1-e^{-r^2/\delta^2(t)} \right).
\end{equation}
\begin{equation}\label{eq:Kambe_p_solution}
p(r,z,t) = p_c(t) -\frac{\rho a^2z^2}{2} -\frac{\rho a^2r^2}{8}+ \frac{\rho M^2} {4\pi^2\delta^2} \left[ \ln2 +E_1(2x) -E_1(x) -\frac{(1-e^{-x})^2}{2x} \right],
\end{equation}
where  $E_1(x)$ is the first-order Exponential Integral defined as
\[
E_1(x) = \int_x^\infty \frac{e^{-s}}{s}\,ds, 
\]
and
\[
\delta^2(t) = \frac{4\nu}{a}(1-e^{-at}), \qquad
x=\frac{r^2}{\delta^2(t)}.
\]
We thus obtain the azimuthal wind field $v(r,t)$, the vertical vorticity field $\zeta(r,t)$ as well as the pressure field $p(r,t)$ consistent with the prescribed wind strain $(u,w)$. 

Kambe's solution (\ref{eq:Kambe_eqn}) contains several interesting features. First, it shows how vorticity grows through stretching. An initial vorticity column will start from the center and expand outward while its vorticity magnitude decreases with time and approaches a fixed amplitude. Second, similar to the Burgers model, Kambe's solution highlights the importance of viscosity, which stops the collapse of the vortex core. The final end stage of this vortex development is a finite-radius steady column, which is approached on the timescale $1/a$.

\section{Shear-forced dynamic model}
Incorporating vertical wind shear (VWS) into a model for tornado dynamics while retaining axisymmetry is a challenging problem. On the one hand, the axisymmetric assumption is essential for simplifying the governing equations in cylindrical coordinates and obtaining tractable analytical solutions. On the other hand, real VWS is intrinsic to tornado development and generally encompasses the complex tornado-environment interaction that cannot be represented as a simple axisymmetric forcing during the entire tornado lifetime. 

In practice, the VWS in tornado development often consists of two components: (i) the 'environmental shear', which acts as an external forcing imposed by the surrounding atmosphere, and (ii) the 'internal vortex shear' arising from the change of tornado vertical structure with height. Because thunderstorm systems that host tornadoes inherently contain environmental VWS throughout their lifecycle and tornadoes most often exhibit some degree of vertical variation in their vortex structure, these two forms of shear are always present.

Between these two VWS components, we note that environmental VWS poses the great theoretical challenge. It can be shown explicitly that incorporating environmental VWS destroys the axisymmetry of tornado structure.\footnote{Using the Galilean translational invariance, one can show that the vorticity equation acquires an azimuthally dependent forcing term in the presence of environmental VWS.} As such, the resulting dynamics become fully three-dimensional and generally require numerical simulations to capture the interaction between the environmental shear and the tornado vortex. From an analytical perspective, there is no straightforward way to incorporate environmental VWS into an axisymmetric tornado model. 

In contrast, the internal vortex shear due to the height dependence of tornado wind structure can be incorporated within the axisymmetric framework as soon as their upright rotational column emerges. If vertical wind $w$ also varies horizontally, the tilting effect will also play a role. The objective of this section is therefore to present a model that captures the dynamical effects of this internally-induced shear forcing. Specifically, we want to prescribe a meridional velocity field $(u,w)$ similar to the strain flow in the Burgers vortex model but it contains an additional radial dependence in the vertical wind $w$. This modification enables the tilting mechanism associated with vortex shear structure to be included within an analytically tractable framework that we wish to present in this Section.
%... instead if when a tornado forms, how would it elove under the shear forcing. Here, the environmental shear still cannot be included in anximmstrical framework. However, its effects on tornnado dynamics can still be felt thru the tiling term in the form of an external forcing. We will paramterize this forcing as $S\partial_r W$, where $S$ is shear-realted magnitude. It is equailvent to separate $v=v_{vortex}(r,t) + v_{env}(z)$.  

\subsection{Vortex-induced shear dynamics}
Consider a general vortex column whose azimuthal wind possesses some degree of vertical variation, i.e., $\partial v/ \partial z \ne 0$. To distinguish this type of vortex structure from the classical Burgers vortex whose azimuthal structure is vertically invariant, we shall refer to such a vortex throughout this section as a \emph{sheared vortex}\footnote{This type of vortex is sometimes referred to as a \emph{baroclinic vortex}. However, we will avoid that terminology here because baroclinicity conventionally includes also thermodynamic effects arising from the solenoidal term.}. Our aim now is to examine the dynamical consequences of the vertical shear generated by such a vortex internal structure variation. Specifically, we look for solutions that show how this internal shear interacts with vertical wind through vortex tilting during the evolution of the sheared vortex.

To this end, we modify the secondary circulation $(u,w)$ in Kambe's model by prescribing the meridional velocity field as follows
\begin{equation}\label{eq:shear_uw_flow}
u=-\frac{a}{2}r, \qquad w=az+br,
\end{equation}
where $a$ characterizes the radial convergence of the inflow, and $b$ represents the rate of change of the vertical velocity with radius. Both parameters have units of inverse time. Since the term $br$ is independent of $z$, the modified velocity field remains incompressible.

In the presence of vortex shear, deriving solutions from the vertical vorticity equation becomes considerably more challenging. This is because one must ensure the simultaneous consistency among the vorticity equation, the vertical variation of the azimuthal wind, and the radial constraint of the azimuthal wind and the vertical vorticity. These constraints make the vorticity formulation analytically unsolvable. Therefore, we will work directly with the azimuthal momentum equation, which governs the evolution of the azimuthal velocity $v$ given by
\begin{equation}\label{eq:azimuthal_wind}
\frac{\partial v}{\partial t} -\frac{ar}{2}\frac{\partial v}{\partial r}
+(az+br)\frac{\partial v}{\partial z} -\frac{a}{2}v = \nu\left(
\frac{\partial^2v}{\partial r^2} +\frac{1}{r}\frac{\partial v}{\partial r}
+\frac{\partial^2v}{\partial z^2} -\frac{v}{r^2} \right).
\end{equation}
The dependence of $v$ on $z$ is important here, so that the VWS effects associated with $\partial v/\partial z$ can be properly taken into account. It should be noted here that even though the equation for angular momentum $\Gamma$ as given by Eq. (\ref{eq:swirl}) may look simpler under the flow settings (\ref{eq:shear_uw_flow}), the radial-vertical coupling remains. Thus, the final solution for $\Gamma$ turns out to be of the same form as for $v$, which cannot be completely expressed in a closed-form as will be shown below. For this reason, we will solve the solution for the azimuthal wind directly from Eq. (\ref{eq:azimuthal_wind}) instead of using the angular momentum equation (\ref{eq:swirl}).    

Consider next the dimensionally consistent initial condition of the following form
\begin{equation}\label{eq:v_init}
v(r,z,0) = A_0r \exp\left(-\frac{r^2}{R_0^2}\right) \cos(kz),
\end{equation} where $[A_0]=\mathrm{s}^{-1}$ is proportional to the initial vortex strength, and $[k]=\mathrm{m}^{-1}$ is the vertical scale of vortex. This choice of the $\cos(kz)$ for the vertical structure of the initial vortex is motivated by the fact the azimuthal wind will be come strongest near surface in the free-slip boundary condition, which decay with height as in Kieu and Zhang (2009,\cite{KieuZhang2009}). 

Follow Kambe's method, we introduce similarity reduction as follows:
\begin{equation}\label{eq:coord_transform}
\xi=re^{at/2}, \qquad \eta=ze^{-at}, \qquad \tau=\frac{e^{at}-1}{a}, \qquad
Q(\tau)=1+a\tau=e^{at},
\end{equation}
and use the complex representation for the azimuthal wind
\begin{equation}\label{eq:complex_transform}
v(r,z,t) = \operatorname{Re}\left\{ e^{at/2}U(\xi,\eta,\tau) \right\}.
\end{equation}
Substitution into Eq.~(\ref{eq:azimuthal_wind}) gives
\begin{equation}\label{eq:transformed_PDE}
U_\tau +b\xi Q^{-5/2}U_\eta = \nu\left( U_{\xi\xi} +\frac{1}{\xi}U_\xi -\frac{U}{\xi^2} \right) +\nu Q^{-3}U_{\eta\eta}.
\end{equation}
The term proportional to $b$, which represent the impact of vortex vertical shear, couples $\xi$ and $\eta$. 

Represent the initial cosine by the real part of $e^{ik\eta}$ and
write
\begin{equation}\label{eq:fourier_mode}
U(\xi,\eta,\tau) = A_0\xi H(\xi,\tau)e^{ik\eta}.
\end{equation}
The complex radial amplitude then satisfies
\begin{equation}\label{eq:H_func}
H_\tau = \nu\left( H_{\xi\xi} +\frac{3}{\xi}H_\xi \right) -\nu k^2Q^{-3}H
-ibk\xi Q^{-5/2}H,
\end{equation}
with
\begin{equation}\label{eq:radial_amplitude}
H(\xi,0) = \exp\left(-\frac{\xi^2}{R_0^2}\right).
\end{equation}
Thus, the physical velocity is
\begin{equation}\label{eq:velocity_from_H}
v(r,z,t) = A_0r e^{at} \operatorname{Re}\left\{ e^{ikze^{-at}}H(\xi,\tau) \right\}.
\end{equation}

We can obtain more explicitly the exact viscous representation for the solution (\ref{eq:velocity_from_H}) by noting that Eq. (\ref{eq:H_func}) can be represented by the Feynman-Kac formula if we write it in the form
\begin{equation}
H_\tau = \mathcal{L}H - V(\xi,\tau)H,
\end{equation}
where the second-order differential operator and the complex-valued potential are given by:
\begin{equation}
\mathcal{L} = \nu\left( \frac{\partial^2}{\partial \xi^2} + \frac{3}{\xi}\frac{\partial}{\partial \xi} \right), \qquad V(\xi,\tau)= \nu k^2Q^{-3} + ibk\xi Q^{-5/2}
\end{equation}
This operator form is exactly the infinitesimal generator of this following diffusion equation:
\begin{equation}\label{eq:diffusion_proc}
d\rho_\sigma = \frac{3\nu}{\rho_\sigma}\,d\sigma +\sqrt{2\nu}\,dB_\sigma,
\end{equation}
where $B_\sigma$ is a standard one-dimensional Brownian motion, and $\rho_\sigma$ denote the four-dimensional radial diffusion process that starts at $\rho_0=\xi$ and satisfies Eq. (\ref{eq:diffusion_proc}). Thus, let's define a function
\begin{equation}\label{eq:s_func}
s(t) = \frac{1-e^{-2at}}{2a}.
\end{equation}
The Feynman--Kac representation of Eq.~(\ref{eq:H_func}) is then given as:
\begin{equation}\label{eq:FeynmanKac}
H(\xi,\tau) = e^{-\nu k^2s(t)} \mathrm{E}_{\xi}\left[
\exp\left(-\frac{\rho_\tau^2}{R_0^2}\right) \times \exp\left( -ibk\int_0^\tau \frac{\rho_\sigma} {\left[1+a(\tau-\sigma)\right]^{5/2}} \,d\sigma \right) \right],
\end{equation}
where the expectation operator $\mathrm{E}_{\xi}[...]$ denotes the average of the enclosed complex quantity over every possible Brownian trajectory from Eq. (\ref{eq:diffusion_proc}) that starts at $\xi$. Intuitively, this expectation operator simply means that we need to evaluate the terminal Gaussian $e^{-\rho_\tau^2/R^2_0}$ for each realized path whose initial value is $\xi$, and then average over all possible paths. There is no closed expression for $\mathrm{E}_{\xi}$, but this can be simply treated as a function that can return us its values for a given set of input arguments. 

Combining Eqs.~(\ref{eq:velocity_from_H}) and (\ref{eq:FeynmanKac}) gives the exact viscous solution

\begin{equation}\label{eq:exact_viscous_sol}
\begin{aligned}
v(r,z,t) =\;& A_0r e^{at}e^{-\nu k^2s(t)} \\
&\times \operatorname{Re}\left\{
e^{ikze^{-at}}
\mathrm{E}_{\xi}\left[
\exp\left(-\frac{\rho_\tau^2}{R_0^2}\right)
\right.\right. \\
&\left.\left.
\qquad{}\times
\exp\left(
-ibk\int_0^\tau
\frac{\rho_\sigma}
{\left[1+a(\tau-\sigma)\right]^{5/2}}
\,d\sigma
\right)
\right]
\right\}.
\end{aligned}
\end{equation}
This solution satisfies Eq.~(\ref{eq:v_init}) and decays with time in the form $\exp[{-\nu k^2 \frac{(1-e^{-2at})}{2a}}]$. Such double exponential-time dependent is similar to that in the solution for tropical cyclone development during the rapid intensification phase obtained in \cite{KieuZhang2009} (but with spin-up instead of decaying as in here). The root for these double exponential time dependence is the nonlinear advection terms, which either spin up or spin down a vortex rapidly. Of course, the timescale of tornadoes differs from that in tropical cyclone development. However, such nested exponential solution could reiterate the role of nonlinear advection in the development of vortex with strong radial convergence.   

Note that the azimuthal wind is real. So, if we write out explicitly $ H=H_R+iH_I$, then
\begin{equation}
\operatorname{Re}\left(e^{ik\eta}H\right) = H_R\cos(k\eta)-H_I\sin(k\eta),
\end{equation}
and Eq.~(\ref{eq:velocity_from_H}) becomes
\begin{equation}\label{eq:Real_v}
v(r,z,t) = A_0r e^{at} \left[ H_R\cos(k\eta)-H_I\sin(k\eta) \right].
\end{equation}
When $b=0$, Eq.~(\ref{eq:FeynmanKac}) reduces to a closed-form as follows:
\begin{equation}\label{eq:inviscid_H}
H(\xi,\tau) = e^{-\nu k^2s(t)} \frac{R_0^4} {\left(R_0^2+4\nu\tau\right)^2}
\exp\left[ -\frac{\xi^2}{R_0^2+4\nu\tau} \right],
\end{equation}
Thus, $H_I=0$ and the vertical structure remains a pure cosine with time. That is, the vortex core will have a fixed vertical structure at all time in the absence of VWS forcing. When $b\ne 0$ (i.e., vertical motion varies horizontally), the term $br\,v_z$ generates a sine component, which represents a radius-dependent phase displacement. This allows for the vortex core structure to vary with time, thus highlighting the role of VWS effects on the development and structure of tornadoes.

\subsection{Effective radius for a sheared vortex}
Before going into some detailed analyses of the role of vortex vertical shear, it is worth to examine several limits for the above solution. First, because the exact finite-time solution for $b\ne0$ is not generally Gaussian, we will now define $R_{\mathrm{eff}}$ from the leading Gaussian radial envelope for $v$ as
\begin{equation}\label{eq:effective_R}
|v| \mathrel{\propto} r\exp\left[-\frac{r^2}{R_{\mathrm{eff}}^2(t)}\right].
\end{equation}
For any viscosity $\nu$, the $b$-dependent coefficient in
Eq.~(\ref{eq:H_func}) becomes asymptotically weak.  The
leading radial profile approaches the Burgers strain--diffusion
balance, so
\begin{equation}\label{eq:vis_effective_R}
\lim_{t\to\infty}R_{\mathrm{eff}}^2(t) = \frac{4\nu}{a},
\qquad \lim_{t\to\infty}R_{\mathrm{eff}}(t) = 2\sqrt{\frac{\nu}{a}}.
\end{equation}
Generically, the corresponding velocity has the asymptotic form
\begin{equation}\label{eq:vis_asymptop_v}
v(r,z,t) \sim C_b\,r \exp\left(-\frac{ar^2}{4\nu}\right)e^{-at},
\qquad t\rightarrow\infty,
\end{equation}
where $C_b$ is a real constant with units of inverse time, which depends on $A_0,R_0,a,b,\nu,$ and $k$, and incorporates the radial--vertical phase mixing.

For the case with no titling effect (i.e., $b=0$), we have
\begin{equation}\label{eq:C_mag}
C_0 = A_0\frac{a^2R_0^4}{16\nu^2} \exp\left(-\frac{\nu k^2}{2a}\right),
\end{equation} 
and the maximum of the leading radial envelope now occurs at a smaller radius of
\begin{equation}\label{eq:vis_max_radi}
r_{\max} \rightarrow \sqrt{\frac{2\nu}{a}} = \frac{R_{\mathrm{eff}}}{\sqrt{2}}.
\end{equation}
At any fixed height, $ze^{-at}\rightarrow0$. Hence, the original vertical cosine becomes locally uniform at long times.

Note further that in the inviscid case, Eq.~(\ref{eq:velocity_from_H}) instead has
\begin{equation}\label{eq:invis_eff_R}
R_{\mathrm{eff}}(t) = R_0e^{-at/2} \rightarrow0.
\end{equation}
which is consistent with the collapse of the vortex core when $\nu=0$ in the Burgers model. In addition, for every fixed $r>0$,
\begin{equation}
\left|v(r,z,t)\right| \le |A_0|r e^{at} \exp\left(-\frac{r^2e^{at}}{R_0^2}\right)
\rightarrow0,
\end{equation}
faster than exponentially. This convergence is not uniform in $r$. That is, the Gaussian envelope collapses toward the axis, and its peak amplitude scales as $e^{at/2}$. Thus, the core width of inviscid vortex will very quickly decay to zero as in the Burgers model.

\subsection{Complete solution for vorticity and pressure}
With the solution for $v$, we can now obtain the solution for the rest of the variables. Recall that for an axisymmetric flow, the vertical component of vorticity is
\begin{equation}\label{eq:def_shear_vort}
\zeta(r,z,t) \equiv\omega_z(r,z,t) = \frac{1}{r}\frac{\partial(rv)}{\partial r}.
\end{equation}
Thus, direct differentiation of the solution (\ref{eq:Real_v}) gives
\begin{equation}\label{eq:exact_shear_vort}
\zeta(r,z,t) = A_0e^{at} \operatorname{Re}\left\{ e^{ikze^{-at}}
\left[ 2H(\xi,\tau)+\xi H_\xi(\xi,\tau) \right] \right\}.
\end{equation}
In terms of the real and imaginary parts of $H$, Eq.~(\ref{eq:exact_shear_vort}) becomes
\begin{equation}\label{eq:exact_shear_vortReal}
\zeta(r,z,t) =A_0e^{at} \Bigg[ \left(2H_R+\xi H_{R,\xi}\right)
\cos\left(kze^{-at}\right) - \left(2H_I+\xi H_{I,\xi}\right) \sin\left(kze^{-at}\right) \Bigg].
\end{equation}
Note that at $t=0$, one has $\xi=r$, $H=\exp(-r^2/R_0^2)$, and the result reduces to
\begin{equation}\label{eq:vort_init}
\zeta(r,z,0) = 2A_0 \left( 1-\frac{r^2}{R_0^2} \right) \exp\left(-\frac{r^2}{R_0^2}\right)\cos(kz).
\end{equation}

For pressure, we will diagnose it  from radial momentum balance, as the vertical momentum equation contain the external forcing $F_z$ that we do not know in advance (see Eq. \ref{eq:momentum}). With the prescribed radial wind in Eq.~(\ref{eq:shear_uw_flow}), the radial momentum balance requires
\begin{equation}\label{eq:BrRadialPressureGradient}
\frac{\partial p}{\partial r} =
\rho\left( \frac{v^2}{r}-\frac{a^2}{4}r \right).
\end{equation}
Substitution of Eq.~(\ref{eq:velocity_from_H}) gives
\begin{equation}
\frac{\partial p}{\partial r}
=
\rho A_0^2e^{2at}r
\left[
\operatorname{Re}
\left(
e^{ikze^{-at}}H(\xi,\tau)
\right)
\right]^2
-\frac{\rho a^2}{4}r.
\label{eq:exact_P_derivative}
\end{equation}
Integrating from the axis to $r$, and using $\chi=\widehat r e^{at/2}$, gives the exact radial-balance pressure:
\begin{equation}\label{eq:exact_P}
\begin{aligned}
p(r,z,t) =\;& p_0(z,t)-\frac{\rho a^2}{8}r^2
\\
&+ \frac{\rho A_0^2e^{at}}{2} \int_0^\xi \chi \left\{ |H(\chi,\tau)|^2
+ \operatorname{Re}\left[ e^{2ikze^{-at}}H^2(\chi,\tau)
\right] \right\} \,d\chi .
\end{aligned}
\end{equation}
Here $p_0(z,t)=p(0,z,t)$ is an arbitrary axis reference pressure, which can be chosen from, e.g., the Burgers-strain axis pressure
\begin{equation}
p_0(z,t) = p_c(t)-\frac{\rho a^2}{2}z^2
\end{equation}
For $b\ne0$ and $\nu>0$, this exact quadrature generally does not reduce to elementary Gaussian functions because $H$ itself is given by the Feynman--Kac expectation.

\subsection{Vortex tilting effects}
Given the above exact solution for the vertical vorticity, it is of interest to compare this solution with Kambe's $\delta$-source vorticity solution in the absence of vortex shear. Recall that the exact shear-induced vertical vorticity associated with
Eq.~(\ref{eq:exact_viscous_sol}) is
\begin{equation}\label{eq:vort_comparison}
\zeta_S(r,z,t) = A_0e^{at} \operatorname{Re}\left\{ e^{ikze^{-at}}
\left[ 2H(\xi,\tau)+\xi H_\xi(\xi,\tau) \right] \right\}.
\end{equation}
while Kambe's $\delta$-source solution, Eq.~(\ref{eq:Kambe_delta_solution}) is
\begin{equation}\label{eq:Kambe_compare}
\zeta_K(r,t) = \frac{M} {\pi R_B^2(1-e^{-at})} \exp\left[ -\frac{r^2}{R_B^2(1-e^{-at})} \right], \qquad R_B^2=\frac{4\nu}{a}.
\end{equation}
The two solutions have the same asymptotic Burgers radial scale $R_B$, but they represent fundamentally different circulation classes. Consider first the long-time limit for which Eq.~(\ref{eq:vis_asymptop_v}) gives:
\begin{equation}
v(r,z,t) \sim C_b r \exp\left(-\frac{r^2}{R_B^2}\right)e^{-at}.
\end{equation}
Applying $\zeta_S=r^{-1}\partial_r(rv)$ therefore gives
\begin{equation}\label{eq:long_time_vort}
\zeta_S(r,z,t) \sim 2C_b e^{-at} \left( 1-\frac{r^2}{R_B^2} \right)
\exp\left(-\frac{r^2}{R_B^2}\right), \qquad t\rightarrow\infty.
\end{equation}
Thus, the new solution approaches the Burgers radial scale but its amplitude now decays as $e^{-at}$. Note also the factor $(1-r^2/R_B^2)$ in (\ref{eq:long_time_vort}), which indicates that the vortex core is shielded inside $R_B$ (i.e., $\zeta_S>0$ for $r < R_B$, assuming the initial vortcity condition $C_b >0$). The sign of the vorticity is reversed for $r> R_B$. This means that the tornadoes will always maintain its definite vorticity in the inner-core region during its decaying due to the existence of vortex shear, once it starts with cyclonic or anticyclonic direction. 

In contrast, Kambe's solution approaches the nonzero steady Burgers vortex:
\begin{equation}
\zeta_K(r,t) \rightarrow \frac{Ma}{4\pi\nu} \exp\left(-\frac{r^2}{R_B^2}\right).
\end{equation}
In terms of the Burgers-operator, Kambe's nonzero-circulation solution consists of the solely non-decaying Gaussian mode for all $r>0$. Thus, the solution remains immune to the destabilizing effects of vortex-induced VWS, since its azimuthal velocity profile is vertically uniform and does not generate higher-order radial modes.

From this perspective, the adverse impact of vortex-induced VWS on tornado development becomes significant only when the tornado core develops vertical variations in its azimuthal wind. Once such a vertical variation emerges, the accompanying horizontal gradients in the vertical velocity produce a baroclinic tornado structure that is intrinsically susceptible to wind shear. The resulting shear continually modifies the vortex core structure, leading to enhanced viscous dissipation and a progressive weakening of the vortex. This mechanism is absent in Kambe's exact vortex solution.

In the far-field limit, note that for every $t>0$, Kambe's vorticity has a positive Gaussian tail:
\begin{equation}
\zeta_K \mathrel{\propto} \exp\left[ -\frac{r^2}{R_B^2(1-e^{-at})} \right]. 
\end{equation}
Its associated azimuthal velocity retains the circulation tail
\begin{equation}
v_K(r,t)\sim\frac{M}{2\pi r}, \qquad r\rightarrow\infty.
\end{equation}

For our new solution, the long-time far-field vorticity follows from
Eq.~(\ref{eq:long_time_vort}):
\begin{equation}
\zeta_S \sim -2C_b e^{-at} \frac{r^2}{R_B^2} \exp\left(-\frac{r^2}{R_B^2}\right),
\qquad r\gg R_B.
\end{equation}
Unlike Kambe's solution, our far-field vorticity solution therefore has the opposite sign to that in the inner core and contains a polynomial prefactor multiplying the Gaussian envelope. This explains why the corresponding azimuthal wind in our solution decays rapidly with radius. At a finite time and for $b\neq0$, the complex radial amplitude $H$ introduces additional radial phase mixing, which may generate further changes in the sign of the vorticity. Thus, in the presence of vertical shear and tilting effects, the far-field vorticity is generally no longer represented by a single, one-signed Gaussian distribution, although its overall envelope remains exponentially localized.

Due to the combination of both cosine and sine functions in (\ref{eq:Real_v}), our solution contains a vertical phase structure that generates radius-dependent cosine--sine mixing before the locally uniform long-time limit is reached. When $w$ varies with radius (i.e., $b\neq0$), the tilting term further modifies the internal structure of the tornado core, shifting the location of the maximum azimuthal wind slantwise upward with height. This feature is not immediately apparent from the analytical solution because it is embedded in the complex radial amplitude $H$. To illustrate this hidden effect of the tilting term and its physical consequences, we present a numerical illustration in the following subsection.  

\subsection{Numerical validation}
Our exact solutions for the shear vortex contains richer information than the original Burger vortex or Kambe's solution. Some of the effects can be analyzed in the limit of long time or far field as presented in the previous subsections. However, some other detailed behaviors cannot be seen directly in the closed-form expression due to the completed Feynman-Kac representation in the expression of $H$ (see \ref{eq:FeynmanKac}).

Before presenting the numerical evaluation of our analytical solution, it is instructive to examine its general qualitative behavior. As noted previously, when $b\neq0$, the vortex structure evolves in time in a restricted but distinct manner. This behavior becomes apparent by expressing the complex Feynman--Kac amplitude as
\begin{equation}\label{eq:AAMHPolarForm}
H(\xi,\tau) = H_R(\xi,\tau)+iH_I(\xi,\tau) =
|H(\xi,\tau)|e^{-i\phi(\xi,\tau)}.
\end{equation}
The azimuthal wind can then be written as
\begin{equation}\label{eq:AAMVerticalStructureVelocity}
v(r,z,t) = A_0re^{at}|H(\xi,\tau)| \cos\left[
kze^{-at}-\phi(\xi,\tau) \right],
\end{equation}
Thus, $H$ contains some time-dependent modulus and complex phase that control the amplitude and radius-dependent displacement of the vertical Fourier mode. In particular, the vertical wavenumber is now given by
\begin{equation}\label{eq:AAMVerticalWavenumber}
k_z(t)=ke^{-at}.
\end{equation}
or equivalently its vertical wavelength is
\begin{equation}\label{eq:AAMVerticalWavelength}
\lambda_z(t) = \frac{2\pi}{k_z(t)} =
\frac{2\pi}{k}e^{at}.
\end{equation}
This shows that each vertical cosine lobe broadens exponentially with time, implying that the vortex core is continuously stretched in the vertical direction any time a localized wind maxima tries to develop near the surface. It should be noted that this stretching occurs even in the absence of the tilting effect ($b=0$), since it originates from the background axisymmetric strain associated with the parameter $a$. The additional effect of a nonzero $b$ is to introduce a radius-dependent phase shift that couples the radial and vertical structures of the vortex, thereby producing a more complex evolution of the tornado core.

While these expressions provide qualitative insight into the solution, its detailed structure can only be obtained through direct evaluation of the Feynman--Kac representation. Accordingly, Table~\ref{tab:scales} summarizes the range of parameters used in the numerical validation of our proposed tornado dynamics model. Among these parameters, the tornado core width scale $R_{\text{eff}}$ is generally not an independent quantity but is determined by the convergence rate $a$ and the kinematic viscosity $\nu$. In the absence of VWS, this relationship follows directly from the Burgers vortex and Kambe's solution, yielding $R_{\text{eff}}^2 = 4\nu/a$. The range of $R_{\text{eff}}$ listed in Table \ref{tab:scales} is therefore consistent with the corresponding range of $a$ only if the eddy kinematic viscosity lies between approximately $1$ and $100~\mathrm{m^2,s^{-1}}$. This range of $\nu$ is within the range of the observed values for unstable convective boundary layers during periods of strong daytime convective mixing or strong wind conditions \cite{Holton2004,ZhangDrenan2012}.
%
% Figure: shear vortex evolution
%
\begin{figure}[ht!]
\centering
\includegraphics[width=16cm]{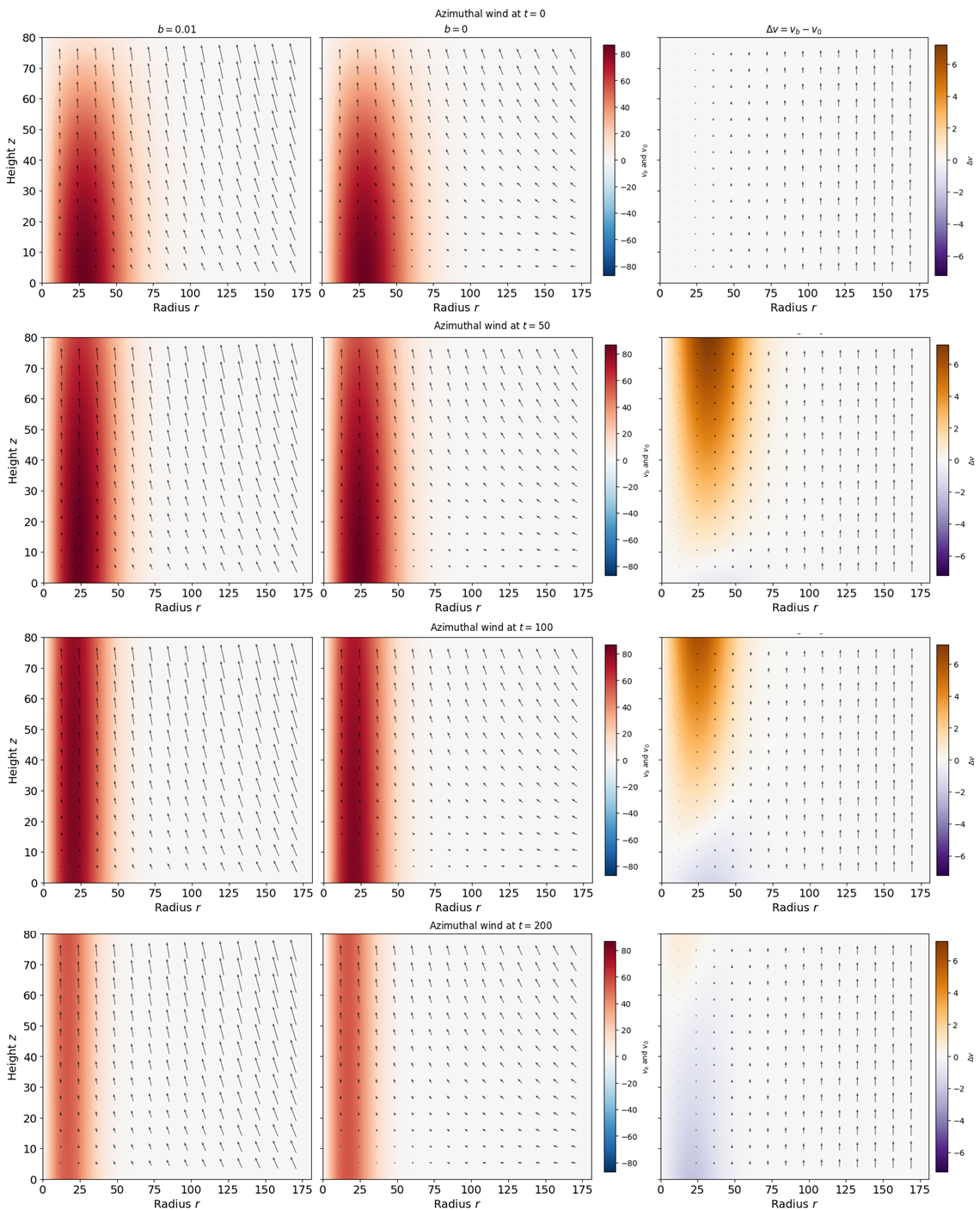}
\caption{An example of the evolution of a shear vortex in the radius-height plane under the increase of vertical wind with radius ($b=0.01$, left panels), no horizontal variation of vertical wind ($b=0$, middle panels), and their respective difference (right panels). Shading denotes the azimuthal wind (unit m s$^{-1}$) and vectors denote the in-plane flow of radial and vertical wind.} 
\label{fig:shear_vortex_rz}
\end{figure}

Figure \ref{fig:shear_vortex_rz} compares the temporal evolution of the vortex obtained from the analytical solutions given by Eqs. (\ref{eq:Real_v}) and (\ref{eq:inviscid_H}), together with the difference between the two solutions in the presence and absence of titling effects for an initial effective vortex core width of $R_{\text{eff}} = 45,\mathrm{m}$. The most noticeable difference is the influence of the tilting effects, which accelerates the upward displacement of the maximum azimuthal wind core. As a result, the tilting term modifies the azimuthal wind structure by more effectively extending and broadening the vortex core aloft.

Along with this upward extension, the vortex core contracts rapidly while the vortex intensity gradually weakens, approaching the asymptotic radius $2\sqrt{\nu/a} = 20,\mathrm{m}$ as deduced from Eq. (\ref{eq:vis_effective_R}). In contrast, in the absence of tilting effects, the vortex core approaches a smaller asymptotic radius as given by Eq. (\ref{eq:vis_max_radi}). The development of vortex shear in the tornado core therefore serves as an indicator of conditions that are unfavorable for tornado maintenance. When embedded in the vertical wind field that varies horizontally, this weakening process is further accelerated by the tilting effects, leading to a more rapid decay of the tornado vortex as indicated by our solution.
  
\begin{table}[ht] \label{tab:scales}
\centering
\caption{Model parameters and scales}
\begin{tabular}{|c|l|l|}
\hline
\textbf{Parameter} & \textbf{Value range (unit)} & \textbf{Remarks} \\
\hline
$a$  &  $10^{-2}-10^{-1}$ (s$^{-1}$)  &  Horizontal convergence rate \\
$b$  &  $10^{-2}-10^{-1}$ (s$^{-1})$ &  The rate of horizontal variation of vertical motion \\
$A_0$  & $10^{-1}-1$ (s$^{-1}$) &  Scaled magnitude of the azimuthal wind profile \\
$\nu$  & $1-10^{2}$ (m$^{2}$s$^{-1}$) & Effective kinematic eddy viscosity \\
$V$  & 40-100 (m s$^{-1}$) & Maximum azimuthal wind (positive for cyclonic vortex) \\
$R_{\text{eff}}$  & 10-300 (m) & Radius of maximum azimuthal wind (core radius) \\
$\Omega$  & $10^{-1}-10$ (s$^{-1}$) & Inner-core rotational angular velocity  \\
\hline
\end{tabular}
\end{table}

\section{Conclusion}
In this note, we extended the Burgers--Kambe axisymmetric vortex framework to account for vertically varying vortex structure and investigated its implications for tornado dynamics. By prescribing an incompressible secondary circulation with radial velocity $u=-ar/2$ and vertical velocity $w=az+br$, where the vertical wind varies linearly with radius, we obtained exact analytical solutions for the tornado wind, pressure, and vorticity fields. The solutions show that the radial variation of the vertical wind (i.e., $b\neq0$) introduces some tilting effects that couple the radial and vertical structures of the vortex core and produce a radius-dependent phase shift. This coupling leads to a rapid reorganization of the tornado core structure. Specifically, the vortex core is quickly stretched vertically, with the maximum azimuthal wind shifted upward and a progressive broadening with height.

Further numerical validation of our solutions confirm that vortex shear helps stretch the tornado core at the rate of $\exp(at)$, even in the absence of the tilting effect. With kinematic viscosity $\nu \ne 0$, the leading radial envelope approaches the Burgers scale $R_B=2\sqrt{\nu/a}$, and the asymptotic vorticity retains its original sign inside the core but reverses sign outside $R_B$. In this regard, our solution captures a shielded, exponentially localized vortex rather than Kambe's one-signed vortex with an algebraic circulation tail. 

Physically, these results reveal how the internal vortex shear can act as a mechanism that enhances viscous weakening and impedes tornado maintenance. Note that such internal vortex shear is expected to emerge after tornado touchdown, because surface friction rapidly modifies the near-surface vortex wind, thus deforming the vortex core and introducing some degree of vertical variations in the tornado core structure. Within the theoretical framework herein, such a structural change can provide an explanation for the transient nature of tornadoes following touchdown.

Of course, our analytical model in this study is intentionally idealized, and its conclusions should not be interpreted as a complete theory of tornado development. A number of drawbkacks should be discussed here. First, our axisymmetric formulation excludes several important physical processes including environmental shear, azimuthal asymmetry, or vortex tilting in real tornado development. Second, the convergence and vertical wind fields are prescribed in our model rather than determined from the governing thermodynamic equations. Therefore, the imposed linear strain field is unbounded outside its intended local region of validity. 

As for the Burges or Kambe's model, additional assumptions such as incompressibility, constant density, constant effective viscosity, or free-slip boundary conditions also simplify the governing dynamics. Accordingly, our solutions should be regarded as an illustration of how tornado-like vortex structures can emerge as an exact solution from the Navier-Stokes equations under idealized forcing. The significance of this approach lies in isolating the dynamical mechanisms responsible for the evolution of the vortex core, rather than in providing a complete description of real tornadoes. Because of this, some behaviors such as the radial phase mixing, upward displacement of the vortex core, or accelerated vortex decay as obtained herein should be viewed as hypotheses that require further validation through observations or fully three-dimensional simulations. Future work should examine the extent to which these mechanisms remain robust when the simplifying assumptions adopted here are progressively relaxed.

%%%%%%%%%%%%%%%%%%%%%%%%%%%%%%%%%%%%%%%%%%%%%%%%%%%%
\section*{Acknowledgments}
This study acknowledges the use of Codex for generating figures based on the Feynman--Kac formula and for validating coordinate transformations used in our solutions.

%\section*{Author contribution} 
%CK perceived the ideas, designed the workflow, analyzed the results, and wrote the draft of this work.

\bibliographystyle{unsrtnat}
\bibliography{references_tornadoes} 
\end{document}